\documentclass[prb,twocolumn,showpacs,noshowkeys,
preprintnumbers,amsmath,amssymb]{revtex4}

\usepackage{graphicx}
\usepackage{dcolumn}
\usepackage{bm}
\usepackage{hyperref}%

\begin{document}

\preprint{}

\title{Multifractality of Hamiltonians with power-law transfer terms}
\author{E. Cuevas}
\email{ecr@um.es}
\homepage{http://bohr.fcu.um.es/miembros/ecr/}
\affiliation{Departamento de F{\'\i}sica, Universidad de Murcia,
E-30071 Murcia, Spain.}

\date{\today}

\begin{abstract}
Finite-size effects in the generalized fractal dimensions $d_q$
are investigated numerically. We concentrate on a one-dimensional
disordered model with long-range random hopping amplitudes in both
the strong- and the weak-coupling regime. At the macroscopic limit,
a linear dependence of $d_q$ on $q$ is found in both regimes for
values of $q \alt 4g^{-1}$, where $g$ is the coupling constant
of the model.
\end{abstract}



\maketitle

\section{Introduction}

Quantum phase transitions in disordered electronic systems remain
one of the central problems in condensed-matter physics. Considerable
attention has recently focused on the critical eigenfunctions, which 
strongly fluctuate near the critical point and thus have multifractal
scaling properties.\cite{HP83,Ja94,FE95,Mi00,CO02} Wave-function statistics
can be characterized through the set of generalized fractal dimensions $d_q$
which are associated with the scaling of the $q$th moment of the wave-function
intensity. A complete knowledge of $d_q$ is equivalent to a complete physical
characterization of the fractal.\cite{HP83}

Metal-insulator transitions (MIT's) depend on the dimensionality
and symmetries of the system and can occur in both the strong
disorder and the weak disorder regime (strong-coupling or weak-coupling
regime, respectively, of the corresponding field-theoretical
description). Each regime is characterized by its respective coupling
strength $g$, which depends on the ratio between diagonal disorder and
the off-diagonal transition matrix elements of the Hamiltonian.\cite{Ef83}

For a multifractal structure of wave functions, the $q$ dependence
of the fractal dimensions $d_q$ is given by the linear relation
\begin{equation}
d_q=d-c_gq \;, \quad q \alt b \;,
\label{dq1}
\end{equation}
where $c_g \ll 1$ and $b$ are parameters related to the coupling
constant of the corresponding model (see below). This equation has
been obtained for different transitions in the weak-coupling limit
using different approaches, which we summarize in the following paragraphs.
Equation (\ref{dq1}) describes \textit{weak multifractality} in the sense
that the fractal dimensions $d_q$ slightly deviate from the system
dimensionality $d$ characteristic of the homogeneously spread metallic
wave functions.

In the context of the theory of Anderson localization, Eq. (\ref{dq1})
was obtained,\cite{FE95} in the leading order in $1/2\pi\nu D \ll 1$,
for two-dimensional (2D) conductors on the basis of a reduced version of
the supersymmetric $\sigma$ model. In this case $c_g=1/2\pi^2\nu \beta D$,
where $\nu$ is the density of states, $D$ the diffusion constant, and
$\beta$ the symmetry parameter.

Using a perturbative $\epsilon$ expansion in one-loop order
for a $d=2+\epsilon$ dimensional system, Eq. (\ref{dq1}) was
found,\cite{We80} in the leading order in $\epsilon \ll 1$,
with $c_g=\epsilon$. This derivation was made for the case
of unbroken time-reversal symmetry (orthogonal universality
class) in the weak-coupling limit.

Reference 8 derived Eq. (\ref{dq1}) with $c_g=g/8\pi$, in the
limit $g \ll 1$ for the power-law random banded matrix model (orthogonal
universality class). This solution was found by mapping the problem
onto a nonlinear $\sigma$ model with nonlocal interaction.

Equation (\ref{dq1}) was also obtained, using the replica trick,\cite{LF94}
for the critical wave function of a 2D Dirac fermion in a random vector
potential (chiral universality class). The $c_g=\Delta_A/\pi$ found
depends on the variance $\Delta_A$ of the Gaussian distributed vector
potential. This result was confirmed recently \cite{MC96,CD01} by
mapping the problem to a Gaussian field theory in an ultrametric space.

Field-theoretical approaches in the context of the integer quantum
Hall transition (unitary universality class) have been addressed recently.
Based on the application of conformal field theory to the critical point
of a 2D Euclidean field theory, Bhaseen \textit{et al.} \cite{BK00} derived
Eq. (\ref{dq1}) with $c_g=2/k$, where $k$ is the inverse coupling constant.

We wish to point out that the existing theories are inherently weak-coupling
results. However, MIT's generically take place at strong disorder, where the
energy scales associated with both disorder and interactions are comparable
to the Fermi energy, unlike to what happens in perturbative $2+\epsilon$
expansion approaches or in effective-field theories. Thus, quantitative
results for $d_q$ are lacking at strong coupling, and finding $d_q$ for
this regime is essential in order to fully understand the MIT's.

The aim of this work was twofold. First, a numerical verification of
the weak-coupling phase result, Eq. (\ref{dq1}). Second, and most importantly,
we were interested in finding the $q$ dependence of $d_q$ in the strong-coupling
regime, which has been left unexplored. Since this case is not accessible by the
previous treatments, the problem can only be addressed by using numerical
calculations.

The paper begins by first giving the model and the methods used for the
calculations in Sec. II. The results for the fractal dimensions in the
weak- and the strong-coupling regimes are presented in Sec. III A and
III B, respectively. Finally, Sec. IV summarizes our findings.

\section{Model and methods}

The disorder-induced MIT is usually investigated for Hamiltonians
with short-range, off-diagonal matrix elements (e.g., the
canonical Anderson model). Other Hamiltonians exhibiting a MIT
in arbitrary dimension $d$ are those that include long-range hopping
terms.\cite{PS98,PS02} The effect of long-range hopping on localization
was originally considered by Anderson \cite{An58} for randomly distributed
impurities in $d$ dimensions with the
$V(\bm{r}-\bm{r'}) \sim |\bm{r}-\bm{r'}|^{-\alpha}$ hopping interaction.
It is known \cite{An58,Le89} that all states are extended for $\alpha < d$,
whereas for $\alpha > d$ the states are localized. Thus, the MIT can be
studied by varying the exponent $\alpha$ at fixed disorder strength.
At the transition line $\alpha=d$, a real-space renormalization group
can be constructed for the distribution of couplings.\cite{Le89,Le99}
These models are the most convenient for studying critical properties
numerically since the exact quantum critical point is known ($\alpha=d$)
and, in addition, they allow the 1D case to be treated, thus using larger
system sizes and reducing the numerical effort. Many real systems of
interest can be described by these Hamiltonians. Among such systems are
optical phonons in disordered dielectric materials coupled by electric
dipole forces,\cite{Yu89} excitations in two-level systems in glasses
interacting via elastic strain,\cite{BL82} magnetic impurities in metals
coupled by an $r^{-3}$ Ruderman-Kittel-Kasuya-Yodida interaction,\cite{CB93}
and impurity quasiparticle states in 2D disordered $d$-wave
superconductors.\cite{BS96}

We will consider here the intensively studied power-law random banded
matrix model (PRBM) \cite{Mi00,ME00,MF96,CG01,CO02,Va02,Cu02,KM97,KT00}.
The corresponding Hamiltonian, which describes a disordered 1D sample
with random long-range hopping, is represented by real symmetric matrices,
whose entries are randomly drawn from a normal distribution with zero mean,
$\left\langle {\cal H}_{ij} \right\rangle =0$, and a variance which
depends on the distance between the lattice sites,
\begin{equation}
\left\langle |{\cal H}_{ij}|^2\right\rangle =\frac{1}{1+(|i-j|/b)^2}
\times\left\{\begin{array}{ll}
                    \tfrac {1}{2}     \ ,\quad & i\neq j\\
                    1 \;\;\;\,\ ,\quad & i=j\;,
       \end{array}\right.
\label{h1dor}
\end{equation}
where the parameter $b$ plays the role of the inverse coupling constant
$b=4g^{-1}$ of the corresponding $\sigma$ model at the center of the
spectral band. The model describes a whole family of critical theories
parametrized by $0<b<\infty$, which determines the critical dimensionless
conductance, in the same way as the dimensionality labels the different
Anderson transitions. In the two limiting cases $b \gg 1$  and $b \ll 1$,
which correspond to the weak and the strong disorder limits, respectively,
some critical properties have been derived analytically.
\cite{Mi00,ME00,KM97,MF96,KT00} More specifically, as already mentioned
in the Introduction, in Ref. 8 the following equation in the limit $b \gg 1$
was derived,
\begin{equation}
d_q=1-\frac {q}{2\pi b}\;, \quad q \ll 2\pi b \;.
\label{dq2}
\end{equation}
The system size ranges between $L=180$ and $L=5400$, and $0.01 \le b \le 12$.
We restrict ourselves to values of $q \alt b$, and consider a small energy
window, containing about 5\% of the states around the center of the band.
The number of random realizations is such that the number of critical states
included for each $L$ is roughly $4\times 10^5$, while, in order to reduce
edge effects, periodic boundary conditions are included. Using methods based
on level statistics, we checked that the normalized nearest level variances
\cite{Cu99} are indeed scale invariant at each critical point studied.

For the computation of $d_q$ we used the standard box-counting procedure,
\cite{Ja94} first dividing the system of $L$ sites into $N_l=L/l$ boxes
of linear size $l$ and determining the box probability of the wave function
in the $i$ box by $p_i(l)=\sum_n |\psi_{kn}|^2$, where the summation is
restricted to sites within that box, and $\psi_{kn}$ denotes the amplitude
of an eigenstate with energy $E_k$ at site $n$. The normalized $q$th moments
of this probability constitute a measure. From this, the mass exponents
$\tau_q(L)$, which encode generalized dimensions $d_q(L)=\tau_q(L)/(q-1)$,
can be obtained,\cite{CJ89}
\begin{equation}
\tau_q(L)=\lim_{\delta \to 0} \frac {\ln \sum_{i=1}^{N_l}
p_i^q(l)}{\ln \delta}\;,\label{tauq}
\end{equation}
where $\delta=l/L$ denotes the ratio of the box sizes and the system size.
It should be made clear that the calculation of $\tau_q(L)$ is suitable only
if the conditions \cite{Ja94}
\begin{equation}
a\ll l < L \ll \xi\ \label{ineq}
\end{equation}
are satisfied, where $\xi$ is the localization or correlation length
and $a$ is the lattice spacing (or any microscopic length scale of
the system). In practice, $\tau_q(L)$ is found by performing a linear
regression of $\ln \sum_{i=1}^{N_l}p_i^q(l)$ with $\ln \delta$ in a finite
interval of $\delta$ (usually $a/L \le \delta \le 1/2$).
In order to properly satisfy the previous conditions (\ref{ineq}) and
(\ref{tauq}), in this work we will compute the derivative 
\begin{equation}
\tau_q(L)=\frac {\textit{d} \ln \sum_{i=1}^{N_l}p_i^q(l)}
{\textit{d}\ln \delta}\biggr|_{\delta=0.1} \;. \label{der}
\end{equation}
First, $\tau_q(L)$ was calculated for different system sizes and then
extrapolated to the macroscopic limit
$\tau_q=\lim_{L \to \infty} \tau_q(L)$.

\section{Results}

\subsection{Weak-coupling regime ($b \gg 1$)}

Using the exact eigenstates of Hamiltonian (\ref{h1dor}) obtained from
numerical diagonalizations, we evaluate, for each value of $q$, $L$,
and $b$, the numerator on the right-hand side of Eq. (\ref{tauq})
for decreasing box sizes, and then calculate $\tau_q(L)$ from the slope
of the graph of the numerator vs $\ln \delta$ at $\delta=0.1$,
Eq. (\ref{der}). Fig. 1 provides an example of the $\ln \delta$
dependence of $\ln \sum_{i=1}^{N_l}p_i^q(l)$ in the weak-coupling
regime ($b=10$) for a system size $L=5400$ and different values of
$q$: $4$ (circles), 5 (triangles), 6 (diamonds), and 7 (squares);
the dotted vertical line corresponds to $\delta=0.1$. The inset shows
the same dependence of the corresponding slopes,
$\textit{d} \ln \sum_{i=1}^{N_l}p_i^q(l)/\textit{d}\ln \delta$.
Note that these derivatives are practically constant in the region shown.
We have checked that the results were practically the same when $\tau_q(L)$
was obtained from the linear fit to
$\ln \sum_{i=1}^{N_l}p_i^q(l)$ vs $\ln \delta$ in the interval
$0.1\alt \delta \alt 0.4$.

\begin{figure}
\begin{center}
\includegraphics[width=6.5cm]{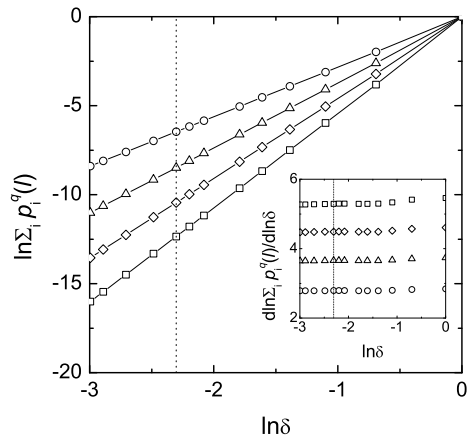}
\caption{\label{fig1} 
The $\ln \delta$ dependence of $\ln \sum_{i=1}^{N_l}p_i^q(l)$ in the
weak-coupling regime ($b=10$) for $L=5400$ and different $q$ values:
$q=4$ (circles), 5 (triangles), 6 (diamonds), and 7 (squares); the
dotted vertical line corresponds to $\delta=0.1$. The inset shows the
same dependence of corresponding slopes,
$\textit{d} \ln \sum_{i=1}^{N_l}p_i^q(l)/\textit{d}\ln \delta$.}
\end{center}
\end{figure}

The size dependence of $\tau_q(L)$ at the critical point of the finite
system for $b=10$ is shown in Fig. 2 for $q=4$. Clearly, significant
finite-size effects are present, and for the larger values of $L$ the
convergence is evident. In order to predict the asymptotic value of
$\tau_q$ from such a plot, a curve of the form
$\tau_q(L)=\tau_q+F_{a_q,c_q}(L)$ was fitted to the data points in this
graph,  where $\tau_q$, $a_q$, and $c_q$ are three fitting parameters and
$F_{a_q,c_q}(L)$ is a function which tends to zero as $L \to \infty$.
Various forms for $F_{a_q,c_q}(L)$ were chosen and tested on the data
plots, including exponential ($\sim e^{-c_q L}$), inverse logarithmic
($\sim 1/\ln c_q L$), and power-law ($\sim L^{-c_q}$) decay with $L$. The
testing involved performing a three-parameter fit on the
$\tau_q(L)$ vs $L$ data plots using the standard Levenberg-Marquardt
method for nonlinear fits. The form of $F_{a_q,c_q}(L)$ eventually chosen
was the power law. The reason for choosing the above form with preference
to the alternative forms was simply that the inverse logarithmic and the
exponential laws did not fit our data properly. In addition, for all values
of $q$ and $b$ we found that the exponent $c_q$ hardly differs from unity.
Thus, we can write
\begin{equation}
\tau_q(L)=\tau_q+a_q/L \;, \label{s2l}
\end{equation}
and keep only two free parameters. The solid line in Fig. 2 is a fit to
Eq. (\ref{s2l}). Note that this equation gives a fairly good fit to the
data. $\tau_q(L)$ is represented on a $1/L$ scale in the inset of Fig. 2
for different values of $q$: $4$ (circles), 5 (triangles), 6 (diamonds),
and 7 (squares). Here, one can better appreciate the linear behavior,
Eq. (\ref{s2l}), of the finite-size corrections to $\tau_q$.

\begin{figure}
\begin{center}
\includegraphics[width=6.5cm]{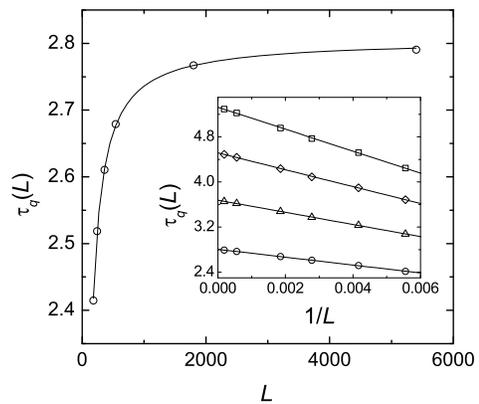}
\caption{\label{fig2} The $L$ dependence of the mass exponent
$\tau_q(L)$ in the weak-coupling regime ($b=10$) for $q=4$
(circles); solid line is a fit to Eq. (\ref{s2l}). The inset
shows $\tau_q(L)$ on a $1/L$ scale for different $q$ values:
$q=4$ (circles), 5 (triangles), 6 (diamonds), and 7 (squares);
the straight lines are linear fits to Eq. (\ref{s2l}).}
\end{center}
\end{figure}

The $q$ dependence of the extrapolated results of $d_q=\tau_q/(q-1)$
in the weak-coupling regime are depicted in the inset of Fig. 3 for
several values of $b$: $6$ (squares), $8$ (triangles), $10$ (diamonds),
and $12$ (circles). The solid lines are fits to the form $d_q=1-c_b q$.
The corresponding $b$-dependent slopes $c_b$ are summarized
and compared with the nonlinear $\sigma$-model estimates $1/2\pi b$
in Fig. 5. The main panel shows the same data $d_q$ as a function of the
rescaled variable $q/2\pi b$, in which it is clearly seen that the data
points collapse fairly well onto a single straight line according to
Eq. (\ref{dq2}) (solid line). To our knowledge, this result provides
the first numerical test of the nonlinear $\sigma$-model prediction
for the PRBM model. A similar linear behavior has been found numerically
for the distribution of the local electric fields at a dielectric resonance
in two dimensions.\cite{JL98}

\begin{figure}
\begin{center}
\includegraphics[width=6.5cm]{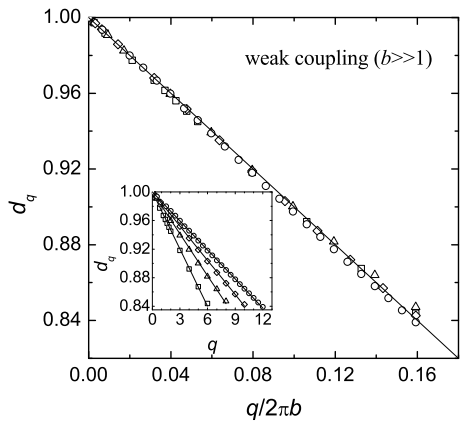}
\caption{\label{fig3} $d_q$ as a function of the rescaled variable
$q/2\pi b$ in the weak-coupling regime ($b \gg 1$) for several $b$
values: $6$ (squares), $8$ (triangles), $10$ (diamonds), and $12$
(circles); the solid line represents the nonlinear $\sigma$-model
estimate, Eq. (\ref{dq2}). The inset shows the same data on the
scale $q$; solid lines are fits to the form $d_q=1-c_b q$.}
\end{center}
\end{figure}

\subsection{Strong-coupling regime ($b \ll 1$)}

We have also calculated $d_q$ for the critical wave functions of 
Hamiltonian (\ref{h1dor}) in the strong-coupling regime for which,
as we already mentioned, there are no analytical predictions for
values of $q \alt b \ll 1$. The only analytical estimate in this
regime was obtained for $q>1/2$, based on a resonance pairs
approximation, in Ref. 22.
$\tau_q \simeq 4b\Gamma(q-\tfrac {1}{2})/\sqrt{\pi}\Gamma(q-1)$
found diverges at $q=1/2$ and for smaller values of $q$ this
approximation completely breaks down.

\begin{figure}
\begin{center}
\includegraphics[width=6.5cm]{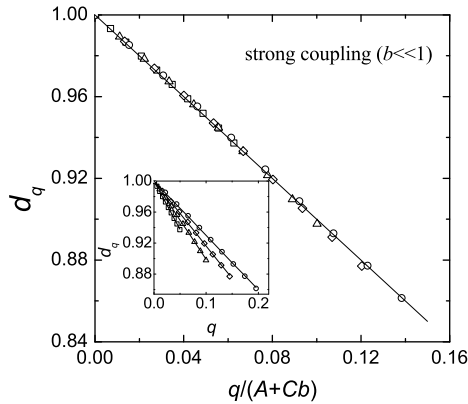}
\caption{\label{fig4} As in Fig. 3, for the strong-coupling regime
($b \ll 1$): $b=0.05$ (squares), $0.1$ (triangles), $0.15$ (diamonds),
and $0.2$ (circles). The solid line corresponds to Eq. (\ref{dq3}).
The straight lines in the inset are fits to the form
$d_q=1-c_b q$.}
\end{center}
\end{figure}

Figure 4 shows the same extrapolated values of $d_q$ as in Fig. 3
in the strong-coupling regime ($b \ll 1$). The values of $b$ reported
are $0.05$ (squares), $0.1$ (triangles), $0.15$ (diamonds), and $0.2$
(circles), and the corresponding slopes $c_b$ are summarized in Fig. 5
(solid circles). As in the weak disorder case, a linear dependence of
$d_q$ on $q$ is found for $q \alt b$. In order to collapse all  data
points onto a single straight line, as was done for the opposite limit
$b \gg 1$, we propose the relation $1/c_b=A+Cb$ for the slopes, where
$A \ne 0$ and $C$ are two fitting parameters (see inset of Fig. 5).
The collapse of the data is evident from the main panel of Fig. 4,
when these data are represented as a function of the rescaled variable
$q/(A+Cb)$. Hence, the $q$ dependence of $d_q$ in the limit $b \ll 1$
can be described by the linear relation
\begin{equation}
d_q=1-\frac {q}{A+C\,b}\;, \quad q \alt b \;.
\label{dq3}
\end{equation}
This equation constitutes the main result of the present paper.
Thus, our results suggest the intriguing possibility that the linear
dependence of $d_q$ on $q$ seems to be a generic law valid in all regimes.
Although this is a conjecture at the present state, we expect that it will
stimulate further analytical work.

\begin{figure}
\begin{center}
\includegraphics[width=6.5cm]{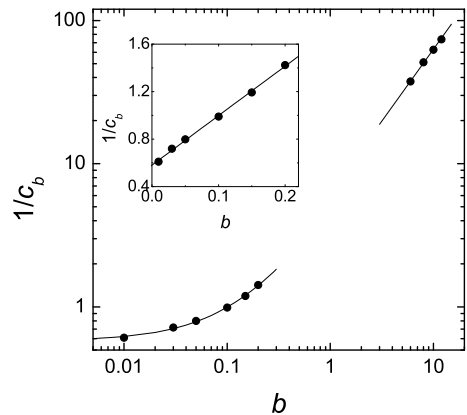}
\caption{\label{fig5} The inverse slope $1/c_b$ (solid circles)
of the fractal dimension as a function of the parameter $b$ of the
PRBM model. The solid lines represent the analytical asymptotic
($b \gg 1$) result $2\pi b$ and the strong-coupling ($b \ll 1$)
result proposed in this work, $A+Cb$. The small-$b$ behavior of
$1/c_b$ is reported in the inset; the straight line is a linear fit
to the form $A+Cb$.}
\end{center}
\end{figure}

The inverse slope $1/c_b$ of $d_q$ is depicted on a log-log plot in Fig. 5
(solid circles) as a function of parameter $b$ of the PRBM model. The inset
shows the small-$b$ dependence of $1/c_b$, from which the linear behavior
on $b$ is evident; the straight line is a linear fit to the proposed form
with parameters $A=0.580 \pm 0.009$ and $C=4.17 \pm 0.08$. The numerical
calculations agree perfectly with the analytical asymptotic in the
weak-coupling regime $b \gg 1$, $1/c_b=2\pi b$, and with the form proposed
in this work for the strong-coupling regime $b \ll 1$, $1/c_b=0.58+4.17b$. 

\section{Summary}

In this paper we have calculated the fractal dimensions $d_q$
of the wave functions for 1D disordered systems with long-range
transfer terms at criticality. The leading finite-size corrections
to $d_q$ decay algebraically with exponents equal to $-1$.
At the infinite-size limit, we have confirmed that, according to
theoretical predictions, $d_q$ is linearly dependent on $q$ in the 
weak-coupling regime. Our calculations strongly suggest that this
behavior is also valid at strong coupling. So, the linear
dependence of $d_q$ on $q$ seems to be a generic law valid in
all disorder regimes.

The question arises as to whether these results are applicable to
other quantum systems, particularly in the 3D Anderson transition
that occurs in the strong disorder domain, and whose similarity with
the PRBM model at $b=0.3$ has been demonstrated for several critical
magnitudes.\cite{CO02}

\begin{acknowledgments}
The author acknowleges the Spanish DGESIC for financial support through
Projects Nos. BFM2000-1059 and BFM2003-03800.
\end{acknowledgments}

\end{document}